\documentclass[%
aip,amsmath,amssymb, reprint, linenumbers
]{revtex4-1}
\usepackage{graphicx}
\usepackage{dcolumn}
\usepackage{bm}
\usepackage[ruled,vlined]{algorithm2e}
\usepackage[utf8]{inputenc}
\usepackage[T1]{fontenc}
\usepackage{mathptmx}
\usepackage{listings}
\usepackage{tikz}
\usepackage{soul}
\usetikzlibrary{arrows}
\usetikzlibrary{trees}
\usepackage{midfloat}
\usepackage{cancel}
\usetikzlibrary{decorations.pathmorphing}
\usetikzlibrary{decorations.markings}
\usetikzlibrary{automata,positioning}
\usepackage{float}
\usepackage{caption}
\usepackage{simplewick}
\usepackage{csquotes}
\usepackage{graphicx}
\usepackage{subcaption}
\usepackage{multirow}
\usepackage{amsmath}
\usepackage{cleveref}
\usepackage{tabularx}
\usepackage{natbib}
\usepackage{braket}
\usepackage{todonotes}

\usepackage{xcolor}
\definecolor{codegreen}{rgb}{0,0.6,0}
\definecolor{codegray}{rgb}{0.5,0.5,0.5}
\definecolor{codepurple}{rgb}{0.58,0,0.82}
\definecolor{backcolour}{rgb}{0.95,0.95,0.92}
\lstdefinestyle{mystyle}{
    backgroundcolor=\color{backcolour},   
    commentstyle=\color{codegreen},
    keywordstyle=\color{magenta},
    numberstyle=\tiny\color{codegray},
    stringstyle=\color{codepurple},
    basicstyle=\ttfamily\footnotesize,
    breakatwhitespace=false,         
    breaklines=true,                 
    captionpos=b,                    
    keepspaces=true,                 
    numbers=left,                    
    numbersep=5pt,                  
    showspaces=false,                
    showstringspaces=false,
    showtabs=false,                  
    tabsize=2
}

\begin{document}

\nolinenumbers
\author{Dibyendu Mondal} \affiliation{Department of Chemistry, Indian Institute of Technology Bombay\\ Powai, Mumbai 400076, India}
\author{Chayan Patra} \affiliation{Department of Chemistry, Indian Institute of Technology Bombay\\ Powai, Mumbai 400076, India}
\author{Dipanjali Halder} \affiliation{Department of Chemistry, Indian Institute of Technology Bombay\\ Powai, Mumbai 400076, India}
\author{Rahul Maitra}
\email{rmaitra@chem.iitb.ac.in} 
\affiliation{Department of Chemistry, Indian Institute of Technology Bombay\\ Powai, Mumbai 400076, India} 
\affiliation{Centre of Excellence in Quantum Information, Computing, Science \& Technology, \\ Indian Institute of Technology Bombay, \\ Powai, Mumbai 400076, India}

\title{Projective Quantum Eigensolver with Generalized Operators}

\begin{abstract}
  Determination of molecular energetics and properties is one of the core challenges in the near-term quantum computing. To this end, hybrid quantum-classical algorithms are preferred for Noisy Intermediate Scale Quantum (NISQ) architectures. The Projective Quantum Eigensolver (PQE) is one such algorithms that optimizes the parameters of the chemistry-inspired unitary coupled cluster (UCC) ansatz using a conventional coupled cluster-like residual minimization. Such a strategy involves the projection of the Schrodinger equation on to linearly independent basis towards the parameter optimization, restricting the ansatz is solely defined in terms of the \textit{excitation} operators. This warrants the inclusion of high-rank operators for strongly correlated systems, leading to increased utilization of quantum resources. In this manuscript, we develop a methodology for determining the generalized operators in terms of a closed form residual equations in the PQE framework that can be efficiently implemented in a quantum computer with manageable quantum resources. Such a strategy requires the removal of the underlying redundancy in high-rank excited determinants, generated due to the presence of the generalized operators in the ansatz, by projecting them on to an internally contracted lower dimensional manifold. With the application on several molecular systems, we have demonstrated our ansatz achieves similar accuracy to the (disentangled) UCC with singles, doubles and triples (SDT) ansatz, while utilizing an order of magnitude fewer quantum gates. Furthermore, when simulated under stochastic Gaussian noise or depolarizing hardware noise, our method shows significantly improved noise resilience compared to the other members of PQE family and the state-of-the-art variational quantum eigensolver.
\end{abstract}

\maketitle

\date{\today}

\section{Introduction}
Recent advancements in quantum information and quantum 
technology have stimulated a great deal of interest in the
development of quantum algorithms to solve 
certain class of classically intractable 
problems. Determination of molecular energetics is one of such
problems due to the exponential
growth of the Hilbert-space\cite{expo,expo1}.
Quantum computers on the other hand, with its principle 
of superposition and entanglement, can handle such problems 
in a tractable manner. Along this line, various classes of
hybrid quantum classical algorithms have gained significant
attention for the determination of molecular energetics on 
the Noisy Intermediate Scale Quantum (NISQ)
architecture\cite{quantum_pro,Trapped_ion}. The 
Variational Quantum Eigensolver (VQE)\cite{Peruzzo_2014} 
is the most popular hybrid quantum classical-algorithm for 
the simulation of many-body systems in the NISQ architectures 
where the associated wavefunction parameters are optimized 
in a classical computer to minimize the energy expectation 
value. While the accuracy critically depends on the 
expressibility of the chosen ansatz, the 
unitary coupled cluster (UCC)\cite{Tilly_2022,ucc_review}
provides a chemistry-inspired parametrization of unitary
operators 
in terms of the anti-hermitian cluster operators that are 
proved to be highly accurate. Given the trainability issue 
of such an ansatz and the disruptive noise profile of the 
NISQ devices, the VQE algorithms often suffer from issues 
like slow convergence due to inaccurate determination of
energies and gradients, and large scale non-linear nature 
of the optimization landscape. These issues are further
amplified under noise as the number of quantum measurements 
required for operator averaging continues to grow. 
Substantial progress has been made in reducing the number 
of measurements required for operator averaging by 
grouping commuting Pauli
operators\cite{op_com,op_com1,op_com2,op_com3,op_com4}, 
leveraging integral factorization techniques\cite{integral} and 
employing efficiently computable components\cite{ham_par,ham_par2,ham_par3} of the operator.
Furthermore, advancements have been achieved in calculating 
analytical gradients on quantum hardware, utilizing techniques 
like the parameter-shift rule\cite{PSR,PSR2} and its 
lower-cost variants\cite{PSR1}. These approaches have made 
gradient-based VQE calculations increasingly more feasible on NISQ devices.

A different approach of hybrid quantum-classical algorithm, the 
Projective Quantum Eigensolver (PQE)\cite{PQE}, has emerged 
as a novel paradigm to solve quantum chemical problems in 
quantum computers by adopting the unique strategy of 
optimizing the parameters through classical coupled 
cluster-like residual minimization. It was previously 
observed that the convergence 
pattern of PQE is more rapid than VQE even in presence of noise. 
Consequently, when started with a fixed structured ansatz, 
VQE warrants more gradient evaluations compared to the 
corresponding residual evaluations required by PQE\cite{PQE}.
Thus within PQE framework, one must aim to 
minimize the number of requisite residual evaluations in order
to minimize the utilization of quantum resources.
Within the disentangled Unitary Coupled Cluster (dUCC) 
framework\cite{dUCC}, PQE determines residual
elements by projecting through each of the singly, doubly, 
or higher-order excited determinants generated by the action of 
cluster operators on the reference Hartree-Fock (HF) state. 
For systems with low to moderate electronic
correlation effects,
dUCC ansatz with singles and doubles can provide a good 
quantitative accuracy with respect to Full 
Configuration Interaction (FCI) energies. However, in systems 
with strong correlation, it becomes 
necessary to include triples or higher-order excitations to 
accurately capture correlation effects. 
This inclusion demands a higher number of quantum 
resources, particularly in terms of the number of 
entangling quantum gates, which poses prohibitive 
challenges for current NISQ devices. Several low-cost 
variants of PQE and Selected PQE (SPQE) have recently 
been formulated
such as-- the CNOT-efficient PQE\cite{magoulas2023cnot} that 
uses qubit-excitation operators\cite{yordanov2020efficient},
methods of moments\cite{kowalski2000method} inspired 
PQE\cite{magoulas2023unitary}
and adiabatically decoupled PQE\cite{patra2024projective, halder2023machine}
as well as its dynamic variant with auxiliary subspace corrections\cite{patra2024towards}
where the principles of adiabatic decoupling\cite{patra2023synergistic,agarawal2021approximate} is adopted
- to mention a few.
Although these variants of PQE can extensively limit 
the resource-requirements such as the CNOT counts
and measurements, none of them theoretically guarantees 
the exclusion of explicit triples and higher-order 
excitation operators (which typically proliferates the 
circuit depth in an uncontrollable manner)
from the ansatz. A potential solution to this issue has been 
addressed within the VQE framework. This involves 
incorporating two-body generalized operators ($\hat{G}$) into the ansatz\cite{uccgsd,uiccsdn}, which can implicitly 
account for higher-order excitation effects through lower rank 
tensor decomposition. However, while optimizing the generalized 
operators in VQE framework is straightforward, their 
incorporation in the PQE imposes significant theoretical
challenges. This is due to the fact that
the action of these two-body generalized operators $\hat{G}$ 
on the reference HF state is nilpotent, giving rise to their
Vacuum Annihilating Condition (VAC): $\hat{G}\ket{\phi_0}=0$,
where $\ket{\phi_0}$ is the HF reference. This characteristic
makes it particularly challenging to derive a direct
closed-form expression for determining residual elements
associated with these generalized operators in the PQE framework.

In this manuscript, we focus on a specific set of two-body
generalized operators known as scattering 
operators\cite{intro_s,iccsdn,anish_st,brueckner_s}, which 
implicitly generate higher-order excitation effects when 
acting on top of low-order excited determinants. The choice of 
such an operator and the structure of the ansatz, although not
the key focus of this manuscript, have been briefly justified 
in the subsequent sections and can also be found elsewhere. 
As the primary objective of this paper, starting from an 
arbitrarily structured ansatz containing such generalized 
operators, we have developed a novel approach to derive a 
closed-form equation for determining their residual elements
that can easily be implemented in quantum computers with 
minimal quantum resources.  
This algorithm will be referred to as
generalized PQE (GPQE)-- an abbreviation we will be using
throughout the manuscript.
Such an undertaking bypasses 
the need of explicit incorporation of higher order 
cluster operators within the traditional PQE framework 
at the cost of prohibitively high quantum resources, but 
are otherwise impossible to exclude in strong correlation 
regime. We also demonstrate that GPQE retains all the 
advantageous features of the traditional PQE, particularly
its superior resilience to hardware noise, over VQE.
With a brief summary of the conventional PQE, its 
selected variant (SPQE) and their associated scaling, we
motivate the readers to the importance of a disentangled ansatz
containing the scatterers. We derive our projective formulation, 
GPQE, in Sec. \ref{Sclosed_proj} where we introduce an 
internally contracted 
projection manifold for reaching to a closed-shell expression
of the scattering residuals. We have demonstrated its 
accuracy and resource efficiency in Sec. \ref{results}
in ideal (noiseless)
as well as in noisy environment, and have convincingly
affirmed its superiority over other members of the 
conventional PQE family as well as VQE. Finally we conclude in
Sec. \ref{concl}.
\begin{figure*}[t]
    \centering
    \includegraphics[width= 14cm, height=7cm]{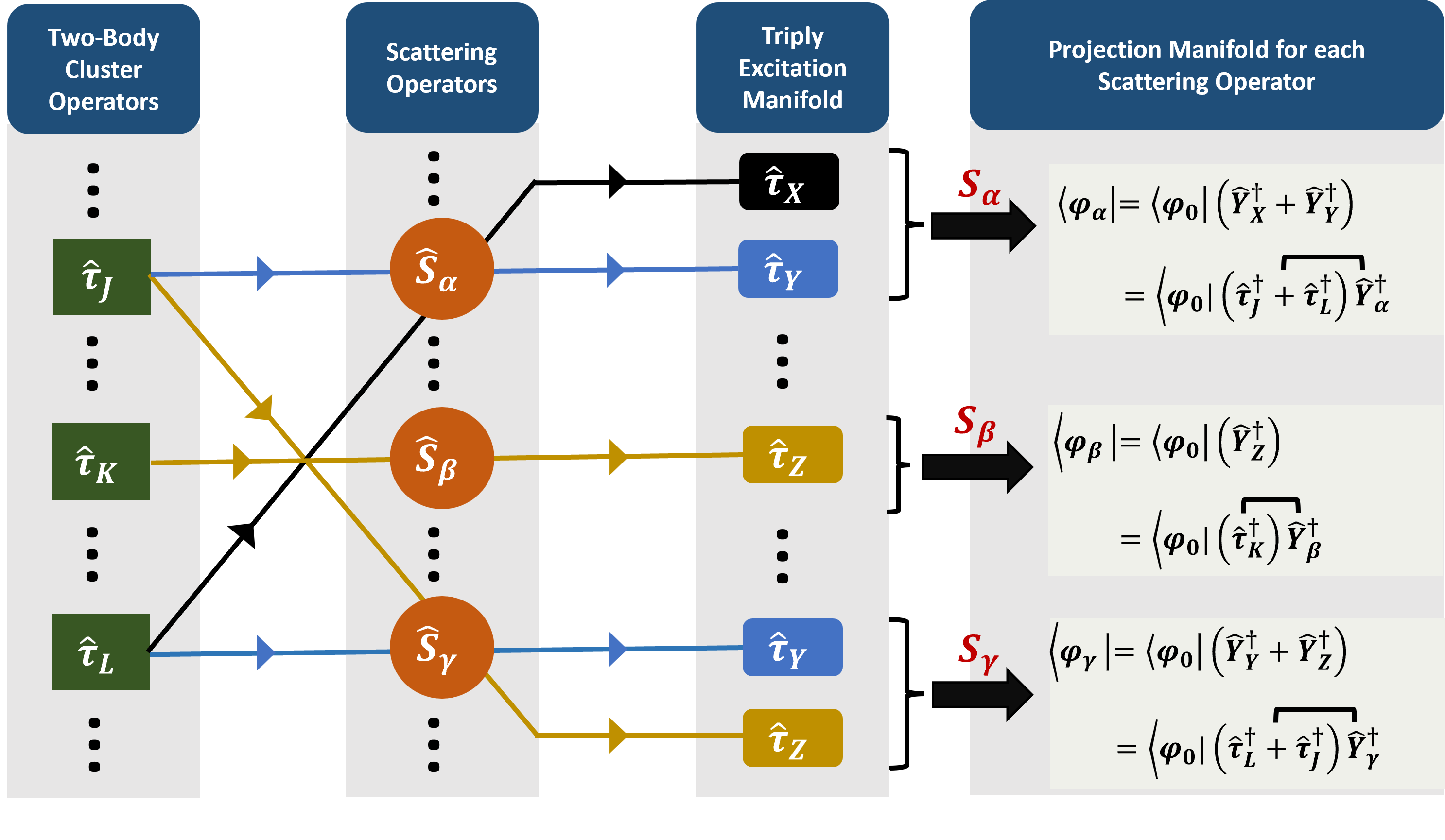}
    \caption{\textbf{Schematic representation of projection manifold for scattering operators}}
    \label{fig:GPQE}
\end{figure*}

\section{Theory}
\subsection{A Brief Summary of Conventional Projective Quantum Eigensolver and its Variants}
\label{recap}
PQE relies on the construction of a parametrized trial wavefunction 
$\ket{\Psi (\boldsymbol{\theta})}$ via the action of a parametrized ansatz
$\hat{U}(\boldsymbol{\theta})$ on the Hartree-Fock (HF) reference state
$(\ket{\phi_0})$: $\ket{\Psi (\boldsymbol{\theta}) }=\hat{U}(\boldsymbol{\theta})\ket{\phi_0}$.
In general one chooses $\hat{U}(\boldsymbol{\theta})$
to be the disentangled unitary coupled cluster (dUCC) ansatz:
\begin{equation} \label{ducc}
    \hat{U}(\boldsymbol{\theta}) = \prod_{\mu} e^{\hat{\kappa}_{\mu}(\theta_{\mu})}
\end{equation}
where, $\hat{\kappa}_{\mu}(\theta_{\mu})=\hat{\tau}_{\mu}(\theta_{\mu})-\hat{\tau}_{\mu}^\dagger(\theta_{\mu})$ is an anti-hermitian cluster operator
with $\hat{\tau}_{\mu}(\theta_{\mu})= \theta_{\mu}\hat{Y}_{\mu}=\theta_{ij...}^{ab...}(\hat{a}^{\dagger}_{a}\hat{a}^{\dagger}_{b}....\hat{a}_{j}\hat{a}_{i})$ and $\theta_{\mu}$
being the associated parameter.
Here, $\mu$ denotes a multi-indexed composite hole-particle \textit{excitation} label
of arbitrary order (single, double, triple or higher order) containing the indices ${i,j,\ldots}$ $({a,b,\ldots})$ that refer to the occupied (unoccupied)
spinorbitals in the HF reference. 
To distinguish between the various order of
excitation operators, we use the notations $\overline{I}, \overline{J}, \overline{K}...$ for single 
excitation operators, $I, J, K...$ for double excitation operators and $X, Y, Z...$ for triple 
excitation operators. This notation is consistently employed throughout the manuscript.
Unlike VQE, PQE adopts a projective
approach leveraging traditional coupled-cluster-like quasi-Newton technique to iteratively optimize
the parameters:
\begin{equation}\label{theta_I} 
\theta_{\mu}^{(n+1)} = \theta_{\mu}^{(n)} + \frac{r_{\mu}^{(n)}}{D_{\mu}}
\end{equation}
Here, $n$ is the iterative step counter and $D_{\mu}$ is the standard M{\o}ller-Plesset
denominator, $D_{\mu}=\epsilon_i+\epsilon_j+...-\epsilon_a-\epsilon_b...$, where the 
$\epsilon_i$ are Hartree-Fock orbital energies. In Eq. \eqref{theta_I}, $r_\mu$ is the 
residual element which can be
constructed as the off-diagonal matrix elements of a similarity transformed 
Hamiltonian 
$\bar{H}=\hat{U}^{\dagger}(\boldsymbol{\theta}) \hat{H} \hat{U}(\boldsymbol{\theta})$
between the excited determinants $\ket{\phi_\mu}$s and the reference 
$\ket{\phi_0}$:
\begin{equation} \label{r_mu}
   r_{\mu}(\boldsymbol{\theta}) = \bra{\phi_{\mu}}\hat{U}^{\dagger}(\boldsymbol{\theta}) \hat{H} \hat{U}(\boldsymbol{\theta})\ket{\phi_o} ; \mu \ne 0.
\end{equation}
Such a projection allows us to have exactly same number of unknown 
parameters as the number of the truncated set of excited determinants.
The residuals can be efficiently calculated using a quantum computer as 
they can be further expressed as a sum of three diagonal quantities:
\begin{equation} \label{r_mu with diagonal quantities}
    r_{\mu} = \bra{\Omega_{\mu}(\frac{\pi}{4})} \bar{H} \ket{\Omega_{\mu}(\frac{\pi}{4})} - \frac{1}{2}E_{\mu} - \frac{1}{2}E_0
\end{equation}
where, $\ket{\Omega_{\mu}(\theta)} = e^{\hat{\kappa}_{\mu}(\theta) } \ket{\phi_0}$ 
, $E_\mu = \bra{\phi_\mu} \bar{H} \ket{\phi_\mu}$
and $E_0 = \bra{\phi_0} \bar{H} \ket{\phi_0}$. This iterative process converges when
the \textit{residual condition} $r_{\mu}\rightarrow 0$ is satisfied.
From the definition of Eq.\eqref{r_mu} it is evident that residuals are nothing but those non-diagonal
elements that form a column (or row) of the matrix representation of $\Bar{H}$ in the many-body determinantal basis.
As a consequence of this structure, the Gershgorin's circle theorem guarantees that
the energy error ($\Delta E$) between the exact ground state energy ($E_{exact}$) and ground state energy obtained by PQE ($E_{PQE}$) is
bounded by $\Delta E = \mid E_{exact} - E_{PQE} \mid \leq \rho = \Sigma_{\mu \neq 0}
r_{\mu}$. Here, $\rho = \Sigma_{\mu \neq 0} r_{\mu}$ is the radius of the Gershgorin's circle and the summation includes only
those residuals for which the residual condition is not enforced. It is crucial to note that the conventional PQE 
necessitates the presence of \textit{only excitation-type operators} 
$(\hat{Y}_\mu)$ in the dUCC ansatz such
that $\hat{Y}_\mu\ket{\phi_0}\rightarrow\ket{\phi_\mu}$. This leads to an unwanted
restriction on the operator pool to construct $\hat{U}(\theta)$: it is 
warranted that high rank excitation operators must be included in 
$\hat{U}(\theta)$ for strongly correlated systems (and such high rank 
connected excitation operators cannot be decomposed into 
lower rank operators), leading to an 
impractical proliferation of circuit depth towards NISQ realization.

The associated dynamic quantum algorithm in the PQE paradigm, known as selected PQE (SPQE)\cite{PQE}, is based upon an \enquote{evolve-and-measure} technique that involves a series of alternating macro- and micro-iteration cycles.
In the macro-iteration steps a low-rank decomposition\cite{berry2019qubitization}
of the Hamiltonian is used to get a \textit{residual state}
via time evolution for a short-period $\Delta t$.
This residual state is subsequently measured to filter a set of \enquote{important} determinants (and excitation operators)
of pre-defined ranks governed by a macro-iteration threshold $\omega$ to construct a compact ansatz. The parameters associated with these specific
important excitation operators are then optimized via PQE micro-iteration cycles. For more theoretical and algorithmic details of SPQE and the
associated resource efficient variants we refer to the paper by Stair \textit{et al.}\cite{PQE} along with some of our recently developed works\cite{patra2024projective,patra2024towards}. 
However, like the parent PQE, its selected variant is also restricted to 
include only the excitation
operators and as such its generalization to incorporate other class of vacuum annihilating operators
require further theoretical development as we described below.

\subsection{Projective Quantum Eigensolver with Generalized Operators: Choice of the Ansatz and Theoretical Challenges}

The idea of expressing a many-electronic wavefunction in terms
of the generalized operators stems from the seminal works by Nooijen\cite{nooijen} and Nakatsuji\cite{nakatsuji}. The subsequent unitary adaptation has motivated its
implementation in the quantum architecture. Following the concepts of 
contracted Schrodinger's equations\cite{CSE}, some of the present authors have 
previously put forward the notion of dual exponential unitary in terms of one and two-body cluster operators and a subset of generalized operators for
improved expressibility of quantum ansatze\cite{uiccsdn,Halder2023}. In general, such an ansatz may
be expressed as: 
\begin{equation} \label{genUni}
    e^{\hat{\lambda}}e^{\hat{\kappa}} = \prod_{\alpha} e^{\hat{\lambda}_{\alpha}(\theta_{\alpha})}\prod_{I} e^{\hat{\kappa}_{I}(\theta_{I})}
\end{equation}
which spans the N-electron Hilbert space via nested commutators: 
\begin{equation} \label{nestedcomm}
e^{\hat{\lambda}}e^{\hat{\kappa}}=e^{\hat{\lambda}+\hat{\kappa}+[\hat{\lambda},\hat{\kappa}]+[\hat{\lambda},[\hat{\lambda},\hat{\kappa}]]+...}
\end{equation}
Structurally, the choice of $\hat{\lambda}=\hat{G}-\hat{G}^{\dagger}$ plays a pivotal 
role in deciphering the expressibility and accuracy of the ansatz with the following two categories: 
\begin{enumerate}
\item Type-1: $\hat{G}$ is chosen to be 
two-body operator with effective hole-particle rank of one and have one 
quasi-orbital destruction.
\item Type-2: $\hat{G}$ is chosen to be 
two-body operator with effective hole-particle rank of zero and have two 
quasi-orbital destruction. 
\end{enumerate}
Note that due to the presence of at least one quasi-orbital
\textit{destruction} operator, both of these set of operators satisfy 
the VAC: $\hat{G}\ket{\phi_0}=0$ (and that is the reason their implementation in the PQE framework is
highly non-trivial irrespective of their position in the ansatz).
While in this manuscript, we do not aim to optimize the circuit
implementation, and thus we will not be making any comparative analysis 
between the two choices stated above; rather, we will start with an 
ansatz of the form expressed by Eq.\eqref{genUni} and would choose 
$\lambda$ to be composed of Type-1 $\hat{G}$ operators such that their rank increasing action
leads to a better span of the $N-$electron Hilbert space\cite{Halder2023}. 
Thus our motivation is to start with an ansatz of structure like that in Eq.\eqref{genUni}
and develop the theoretical methodology to solve within PQE framework so that 
the advantages of PQE is retained while at the same time a desired accuracy may be
achieved with less quantum resources.
Here we start with some further discussions about
$\hat{G}$ (of Type-1) that will motivate us toward our development.
We will refer to these specific operators ($\hat{G}$ of Type-1) as \textit{scatterer}
and denote it as $\hat{S}$.

The scatterers are a class of generalized two-body operators with one 
quasi-orbital destruction operator and depending on whether such destruction 
operators are hole or particle type, we designate them as $\hat{S}_h$ and $\hat{S}_p$
respectively: $\hat{S}=\hat{S}_h+\hat{S}_p$
\begin{eqnarray}
    \hat{S}_h =\frac{1}{2}\theta_{\alpha}\hat{Y}_\alpha=\frac{1}{2}\theta_{i,j}^{a,m}\hat{a}^{\dagger}_{a}\hat{a}^{\dagger}_{m}\hat{a}_{j}\hat{a}_{i} \\ \nonumber 
    \hat{S}_p =\frac{1}{2}\theta_{\beta}\hat{Y}_{\beta}=\frac{1}{2}\theta_{i,e}^{a,b}\hat{a}^{\dagger}_{a}\hat{a}^{\dagger}_{b}\hat{a}_{e}\hat{a}_{i}
\end{eqnarray}
Here $\alpha, \beta$ are the composite indices associated with the orbital indices
of $\hat{S}$. 
The indices $m$ and $e$ refer to a set of occupied (hole) and 
unoccupied (particle) spinorbitals in the reference HF state, and they together 
form a contractible set of orbitals (CSOs). The essential difference 
between the cluster and scattering operators lies in their action on the 
reference determinant: while the action of $\hat{\tau}$ on the reference HF determinant 
generates an excited determinant, the action of scatterer leads to annihilation of the reference leading to the VAC: $\hat{S}\ket{\phi_0}=0$. 
The non-commutativity between the cluster operators and scatterers is
exploited to simulate connected higher order excitations:
\begin{equation}
    \sum_m \hat{S}_{k,j}^{c,m}\hat{\tau}_{i,m}^{a,b}\rightarrow\hat{\tau}_{i,j,k}^{a,b,c}; \sum_e \hat{S}_{k,e}^{c,b}\hat{\tau}_{i,j}^{a,e}\rightarrow\hat{\tau}_{i,j,k}^{a,b,c}
\end{equation}
As the scattering operators have an effective hole-particle excitation rank 
one, each such contraction between $\hat{\tau}$ and $\hat{S}$ increases the excitation 
rank by one: ${\contraction{}{\hat{S}}{}{\hat{\tau_2}}
\hat{S} \hat{\tau_2}}\rightarrow \hat{\tau_3}$,
$\contraction[2ex]{}{\hat{S}}{}{\contraction{}{\hat{S}}{}{\hat{\tau}_{2}} \hat{S} \hat{\tau}_2} \hat{S} \contraction{}{\hat{S}}{}{\hat{\tau}_{2}} \hat{S} \hat{\tau}_{2}
\rightarrow \hat{\tau}_4$... . 
However, such a lower rank decomposition of the higher rank excitations may
often lead to redundant description (\textit{vide infra}). This condition, 
coupled with the associated VAC makes a projective formulation to determine 
the corresponding scatterer rotations highly nontrivial.

\begin{figure*}[t!]
\hspace*{-0.8cm}
    \centering
    \includegraphics[width= 17cm, height=9.0cm]{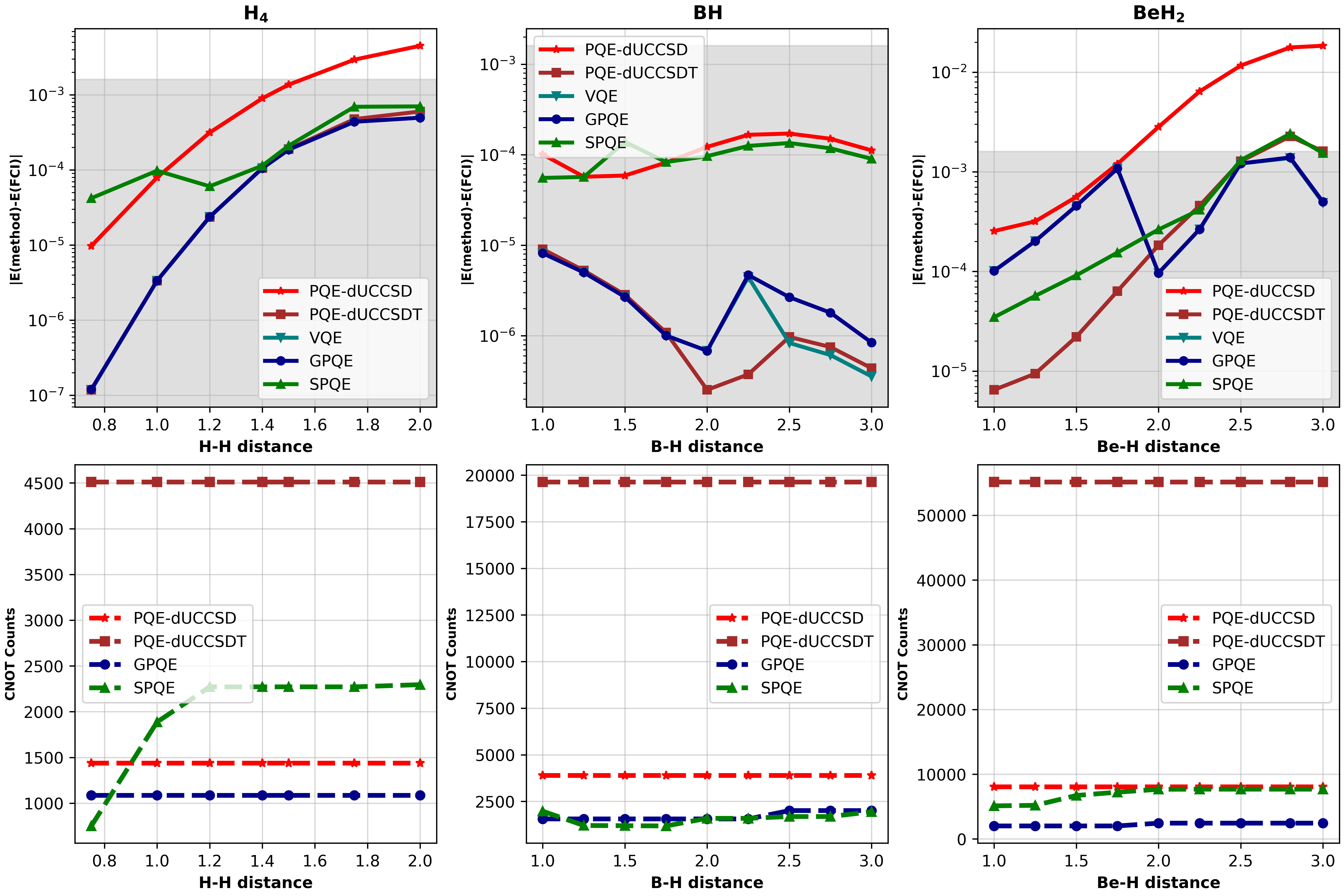}
    \caption{\textbf{Energy difference profile (from FCI, in logarithm scale, along the first row) and CNOT gate count (along the second row) for our GPQE, PQE-dUCCSD, PQE-dUCCSDT, SPQE (with SDT pool) and VQE over the
    potential energy surface. The shaded region indicates the chemical accuracy. Please note that the same ansatz is used for both GPQE and VQE and their 
    results are nearly identical.}}
    \label{fig:surface}
\end{figure*}

\subsection{Towards the Construction of Residue Equation for Scattering Operators}
\label{Sclosed_proj}

The discussion in the preceding section suggests that the scattering operators 
cannot be determined directly by a projective formulation due to the associated 
VAC. However, noting these scatterers have non-zero action on certain doubly 
excited determinants leading to the three body excited determinants, in
principle, such operators may be determined by computing the
matrix elements of an effective Hamiltonian operator between a triply and 
certain doubly excited determinant via Hadamard tests.
However, projection by triply excited determinants often leads to
over-determinedness of the scattering amplitudes, as we explain below.

When the three-body cluster operators are explicitly 
included in the ansatz, the number of projections by 
triply excited determinants are precisely equal to the number 
of unknown three-body parameters. However, when such a
triple excitation is decomposed into a tensor product
of two two-body operators (like scatterers and the cluster 
operators), such one-to-one mapping between the number of
triply excited determinants ($~O(n_o^3n_v^3)$) and the number 
of unknown parameters ($~O(n_o^2n_o^{CSO}n_v + n_on_v^{CSO}n_v^2)$) no longer exists. This arises from the fact
that the action of the scattering operators on various two-body
excited determinants may lead to the redundant generation 
of triply excited determinants: 
each such three-body cluster operator may be generated by more than one combinations of $\hat{Y}_\alpha$ and 
$\hat{Y}_I$. Also, this implies that each element $\hat{Y}_\alpha$ may get coupled with different
$\hat{Y}_I$'s (depending on the commonality of CSO) to generate different $\hat{Y}_X$:
\begin{equation}
\label{sumoverI}
\sum_I (\contraction{}{\hat{Y}_{\alpha}}{}{\hat{Y}_I}\hat{Y}_{\alpha}\hat{Y}_I)\ket{\phi_0}=\sum_X \hat{Y}_X\ket{\phi_0}\rightarrow\sum_X \ket{\phi_X}    
\end{equation}
where we have deliberately put the explicit summation over $I$ for clarity. 
This implies that the structure of $\hat{Y}_\alpha$ is subsumed in each 
such effective three-body excitation operator $\hat{Y}_X$. 
Furthermore, the sum over the 
triply excited determinants arises due to the internal 
summation over various $\hat{Y}_I$'s, 
each sharing a common CSO with $\hat{Y}_\alpha$. 
Thus the direct determination the effective hamiltonian 
matrix elements between the triply and doubly excited 
determinants is theoretically wrong as they have unequal number of
such matrix elements as the number of unknowns, and 
one needs to judiciously contract the three-body projection 
manifold such that the number of projections equals the 
number of unknown parameters.

In order to bypass the redundancy, we propose 
a basis transformation to a lower dimensional manifold
in which a contracted basis vector
$\ket{\phi_{\alpha}}$ is obtained by transforming the $\ket{\phi_X}$ basis via a rectangular matrix such that
\begin{equation}\label{basis rotation eq}
    \ket{\phi_\alpha} = \sum_{I}\theta_I \Big({\contraction{}{\hat{Y}_{\alpha}}{}{\hat{Y}_{I}}\hat{Y}_{\alpha} \hat{Y}_{I}} \Big) \ket{\phi_{0}}= \sum_{X} C_{\alpha X} \ket{\phi_X}
\end{equation}
where, $C_{\alpha X}$ is a rectangular transformation matrix.
Here in Eq.\eqref{basis rotation eq}, for a particular $\alpha$ 
only those elements in the transformation matrix $C_{\alpha X}$
are non-zero for which $\hat{Y}_\alpha$ is structurally a 
subpart of $\hat{Y}_X$. This implies that the associated
$\ket{\phi_X}$ may be obtained via tensor contraction between 
$\hat{Y}_\alpha$ and the corresponding $\hat{Y}_I$s. In this 
sense, it is a dimensionality reduction approach via
a transformation to a lower dimensional basis:
\begin{equation}
    \{\ket{\phi_S}, \ket{\phi_D}, \ket{\phi_X}\} \rightarrow \{ \ket{\phi_S}, \ket{\phi_D}, \ket{\phi_\alpha}\}
\end{equation}
such that $dim(\ket{\phi_\alpha})$ is $~O(n_o^2n_o^{CSO}n_v + n_on_v^{CSO}n_v^2)$ which is $<<dim(\ket{\phi_X})$ having 
$~O(n_o^3n_v^3)$ elements.

With the knowledge of $\theta_I$'s obtained via Eq.\eqref{theta_I} with an ansatz 
$\hat{U}(\theta)=\prod_{\alpha}e^{\hat{\sigma}_\alpha(\theta_\alpha)} \prod_I e^{\hat{\kappa}_I(\theta_I)}$, where $\hat{\sigma}_{\alpha}=\hat{S}_{\alpha}-\hat{S}_{\alpha}^{\dagger}$,
we construct
a residue for $\hat{S}_\alpha$ by explicitly taking the matrix element of
$\bar{H}=\hat{U}^{\dagger}(\theta)H\hat{U}(\theta)$ between the reference 
$\ket{\phi_0}$ and a set of contracted triply excited determinants $\sum_I \bra{\phi_0}(\contraction{}{\hat{Y}_{I}^\dagger}{}{\hat{Y}_\alpha^\dagger}\hat{Y}_{I}^{\dagger}\hat{Y}_{\alpha}^{\dagger})\theta_I$ which effectively spans the transformed 
lower dimensional manifold $\{\ket{\phi_\alpha}\}$.
\begin{equation}\label{S_res1}
    r_{\alpha}(\boldsymbol{\theta}) = \sum_{I}\theta_I\bra{\phi_{0}}\Big({\contraction{}{\hat{Y}_{I}^\dagger}{}{\hat{Y}_{\alpha}^\dagger}\hat{Y}_{I}^\dagger \hat{Y}_{\alpha}^\dagger} \Big)\hat{U}^{\dagger}(\boldsymbol{\theta}) \hat{H} \hat{U}(\boldsymbol{\theta})\ket{\phi_0}
\end{equation}
The summation over the index $I$ is equivalent
to a restricted summation over the resultant three-body
excitations $X$ as Eq. \ref{sumoverI}.
Hence in this lower dimensional basis, residuals $r_\alpha$ are basically the
non-diagonal column (or row) elements of the matrix representation of the similarity transformed Hamiltonian $\bar{H}=\hat{U}^\dagger(\theta) \hat{H} \hat{U}(\theta)$.

Following the conventional approach, the above expression
can be further broken down into the sums of three diagonal 
terms which can efficiently be determined in quantum computers:
\begin{equation} \label{r_S}
    r_{\alpha}(\theta) = \sum_{I:({\contraction{}{I}{}{\alpha}
I \alpha}\rightarrow X)} \theta_{I}\Big(\bra{\Omega_{X}(\frac{\pi}{4})} \bar{H} \ket{\Omega_{X}(\frac{\pi}{4})} - \frac{1}{2}E_{X} - \frac{1}{2}E_0\Big)
\end{equation}
where the terms have their usual meaning as defined in Sec. \ref{recap}.
The residue corresponding to $\hat{S}_\alpha$ now has a particularly simple 
structure: it is structurally akin to the residue for cluster operators,
however, each such residual is explicitly summed over all the three-body excitation terms that has the common $\hat{Y}_\alpha$ subsumed in it. The presence of the associated $\theta_I$'s ensures that 
the \enquote{contracted} projection manifold is weighted by all the 
underlying two-body determinants on which the particular 
$\hat{Y}_\alpha$ has a non-vanishing action.

Following the construction of the residue via its projection
against the contracted excited determinants, the scattering amplitudes
may be iteratively updated using the usual quasi-Newton scheme:
\begin{equation} 
\label{s_update}
\theta_{\alpha}^{(n+1)} = \theta_{\alpha}^{(n)} + \frac{r_{\alpha}^{(n)}(\theta)}{D_{\alpha}}
\end{equation}
where $D_{\alpha}$ denotes the local M{\o}ller-Plesset
denominator for a $\hat{S}_\alpha$. Eq. \eqref{theta_I}
and \eqref{s_update} together generate a set of coupled equations
that are iteratively solved to obtain the converged 
parameters. With this, we move to the next section where we 
discuss the efficiency of GPQE by comparing against allied
PQE based methods and VQE under ideal and noisy environment.

\section{Results and Discussions}
\label{results}
\begin{figure*}[t]
    \centering
    \includegraphics[width= 18cm, height=5.5cm]{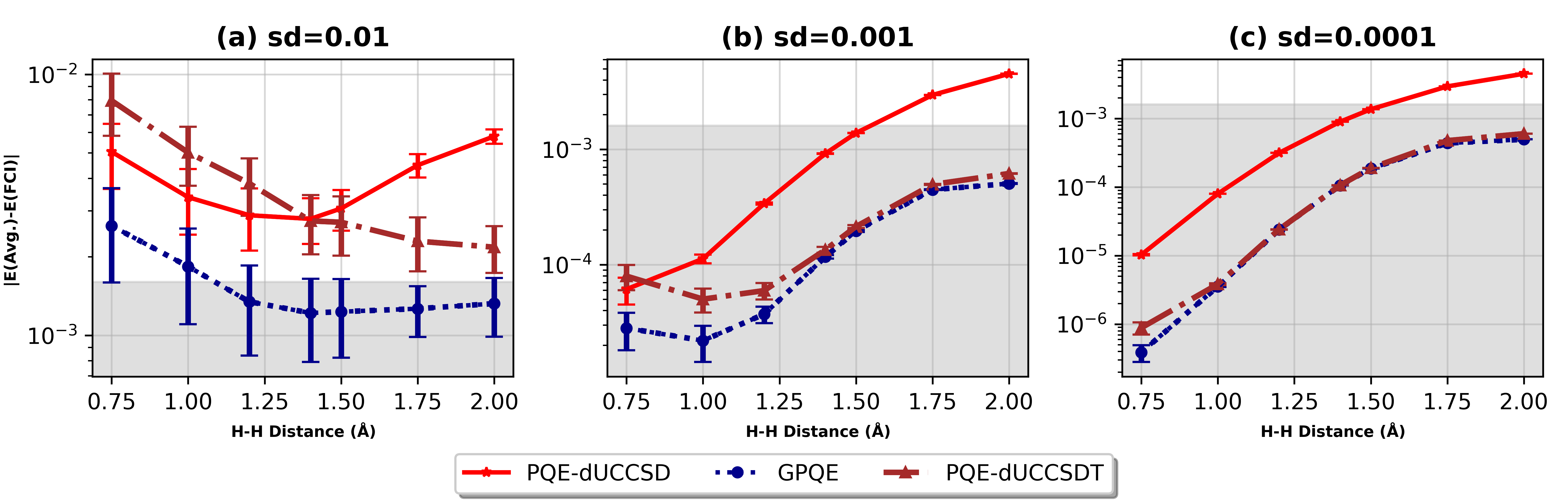}
    \caption{\textbf{Accuracy in energy with respect to FCI is plotted at several internuclear geometries for linear 
    $H_4$ under Gaussian noise model characterized by standard deviation (sd) (a) $10^{-2}$, (b) $10^{-3}$ and (c) $10^{-4}$. At each geometry, the energy is averaged over 100 independent samples under Gaussian noise and error
    bars denote the standard deviation.}}
    \label{fig:noise}
\end{figure*}
\subsection{General Considerations}
In this work, all implementations were carried out using the Qiskit-Nature interface\cite{qiskit_nature}, which imports 
one- and two-body integrals as well as orbital energies from PySCF\cite{pyscf}. We employed the STO-3G basis set\cite{sto} 
for all the systems with direct spinorbital to qubit mapping. 
The Jordan-Wigner encoding was used to convert second-quantized
fermionic operators into qubit operators\cite{JWT}. To 
accelerate the optimization trajectory in PQE framework, the Direct Inversion of 
the Iterative Subspace (DIIS) was applied\cite{diis}. We have used
Broyden-Fletcher-Goldfarb-Shannon (BFGS)\cite{B,F,G,S} algorithm (as implemented in SCIPY\cite{scipy} 
library) for VQE optimization.
For SPQE
calculations we used QForte\cite{stair2021qforte} with 
time-evolution parameter $\Delta t=0.001$ and
macro-iteration threshold $\omega=0.01$, while keeping the maximum excitation rank to order three (i.e. singles, doubles and triples). In order to keep the gate depth to the 
minimum for our formulation, we selected only certain cluster operators 
and scatterers as described below. Please note that for an unbiased comparison, both 
GPQE and the VQE simulations (wherever applicable) employed the same decomposed 
ansatz.

\subsection{Selection of the dominant operators}

To work with the most compact form of the ansatz within the partially disentangled
unitary framework, one may judiciously choose certain operators while the seemingly 
unimportant ones are neglected altogether. For all the two-body operators ($\hat{\tau}_I$'s and 
$\hat{S}_\alpha$'s), we resort to the corresponding first order perturbative measures to
prune the list of the operators: only those operators are chosen for which the
absolute magnitude of their first order estimate\cite{compact} is greater than $10^{-5}$.
The singles are chosen based on the second order perturbative estimate since they appear 
in the second order of many-body perturbation theory for the first time. This implies
that only those singles are chosen where
\begin{equation}\label{sin_select}
    \Big|\mbox{Amp}\big(\frac{({\contraction{}{V^{\dagger}_\alpha}{}{V_I}
V^{\dagger}_\alpha V_I})_{\overline{K}}}{D_{\overline{K}}D_I}\big)\Big|> 10^{-6}
\end{equation}
where $V$ denotes the two-electron integrals. Furthermore, with the pruned set of
scatterers, only those are retained for which both the quasi-hole or quasi-particle
creation operators (for $\hat{S}_h$ and $\hat{S}_p$, respectively) carry paired orbital labels.
Additionally, as demanded by the formulation, the quasi-hole and quasi-particle
destruction operators are restricted to only certain orbitals spanning the CSOs.

\subsection{Accuracy and CNOT gate counts over the Potential Energy Surface: Simulation under noiseless environment}

In this section, we study the performance and resource 
efficiency of our method with noiseless simulator, and compared and
contrasted our results with PQE-dUCCSD, PQE-dUCCSDT, SPQE (SDT pool). The
absolute accuracy of all these 
methods are measured against Full Configuration Interaction (FCI).

The simultaneous symmetric stretching of all the bonds in linear 
$H_4$ chain is a well-studied model system for studying electronic 
strong-correlation behaviour in quantum many-body theories. In the 
STO-3G basis, $H_4$ contains 4 electrons in 8 spinorbitals, which 
are directly mapped onto 8 qubits. For our study we varied the 
$H-H$ bond distance ($R_{H-H}$) from 0.75\AA to 2.0\AA. 
For all geometries, 
we included the HOMO and LUMO spinorbitals in CSOs towards the simulation
of triples. As shown in Fig.\ref{fig:surface}, GPQE 
demonstrates comparable, and in some cases superior 
accuracy to the PQE-dUCCSDT ansatz, while requiring less than one-third 
the number of CNOT gates.

The single bond dissociation of $B-H$ is our next test set. 
With frozen $1s$ orbital of $B$, it renders to be a system with 
4 electrons in 10 spinorbitals. It was observed that for 
$B-H$ bond lengths ranging from 1.0\AA to 2.25\AA, there is no 
low-lying particle orbital through which the scatterer may contract 
with a selected cluster operator. This essentially means that for 
these geometries, $\hat{S}_p$ has no role to play and the CSO is 
constituted by HOMO and (HOMO-1). Beyond $R_{B-H}=2.25\AA$, 
the CSO is composed of HOMO and LUMO. Fig.\ref{fig:surface} (second
column) demonstrates the energy profile of GPQE that 
closely matches with the PQE-dUCCSDT across the PES while the former 
requiring almost an order fewer CNOT gates than the latter.
SPQE with SDT pool utilizes comparable number of CNOT gates with
our ansatz, however, at the cost of compromised the accuracy.

The next system in our study involve the symmetric stretching of 
$Be-H$ bonds in linear $BeH_2$. With the $1s$ orbital of $Be$ frozen, 
it has 4 electrons in 12 spinorbitals. For symmetry considerations,
the CSO contains HOMO and (HOMO-1) up to $R_{Be-H}= 1.75$\AA, while
it involves HOMO and LUMO beyond this point.
The energy accuracy obtained from our ansatz is significantly 
better than dUCCSD across the potential energy surface, even
while using less than one-third of the CNOT gates than the latter.
As shown in the Fig.\ref{fig:surface} (third column), for the 
initial few points, SPQE (with the SDT pool) and dUCCSDT ansatz
provide somewhat better accuracy than GPQE.
However, the corresponding CNOT counts for SPQE and PQE-dUCCSDT
are almost double and an order of 
magnitude higher compared to GPQE, respectively.
Interestingly, in the strongly correlated regions, the accuracy 
of GPQE surpasses both SPQE and PQE-dUCCSDT, as demonstrated in 
the figure.

At this stage, one may note that in the present approach, 
we have worked with an ansatz of the form given in Eq. \ref{genUni} where all the scatterers act after the action of
all the cluster operators. In both the cases, the operators are 
taken in lexical ordering where the one body cluster operators
act first on the reference, followed by the two body cluster
operators and finally by the scatterers. However, one may also
choose to work with a differently ordered ansatz where the 
appearance of the scatterers and the cluster operators are 
interwoven\cite{COMPASS,compact,rbm}. We point out that in such cases, our 
solution strategy for both set of the parameters remains 
unchanged as long as each of the scatterers and cluster 
operators appear only once. In a future endeavour, we will 
be extending our methodology to solve for cases where 
the operators appear more than once.

\subsection{Simulation with a Gaussian Noise Model: Superior Accuracy over dUCCSDT-PQE}
The demonstrated accuracy of our method compared to PQE-dUCCSDT
under ideal noiseless environment
(with the former requiring at least 1-2 orders of magnitude less 
quantum resources) corroborates the balanced treatment of correlation
over the potential energy profiles. However, with the present devices
plagued by different sources of hardware noise, the improved
performance of our method still needs to be validated under
actual or simulated hardware noise. With sufficiently large 
number of experiments performed in NISQ devices, the measurement 
outcomes of an observable may be faithfully sampled by a Gaussian
distribution around its optimal value.
Toward this, for each set of optimal parameters generated under
the ideal noiseless conditions, we simulate 100 sets of noisy parameter 
samples by applying a Gaussian distribution model\cite{Gau,COMPASS}.
Centered around the optimized PQE parameters ($\theta_{opt}$), these
noisy parameters ($\theta_{noisy}$) are distributed according to 
a Gaussian model with a definite standard deviation ($sd$).
\begin{equation}
    \theta_{noisy}\leftarrow exp\Big(-\frac{\theta-\theta_{opt}}{2(sd)^2}\Big)
\end{equation}
One note that this noise model may not fully account for more 
complex or device-specific errors such as decoherence. In our
study, we have considered three different standard deviations, 
$sd = 10^{-2}, 10^{-3}$ and $10^{-4}$ to represent three different
noise strengths and tested the accuracy of various ansatze over 
the linear $H_4$ potential energy profile. Fig. \ref{fig:noise}
conspicuously demonstrates better noise-resilience of GPQE 
compared to PQE-dUCCSD and PQE-dUCCSDT ansatze. Even under 
significant noise strength of $sd=10^{-2}$, the energy 
calculated using GPQE lies within chemical accuracy 
while PQE-dUCCSD and PQE-dUCCSDT deviates substantially, justifying 
its superior noise-resilience within PQE family of methods.

\begin{figure}[h!]
    \includegraphics[width=8 cm,height=15 cm]{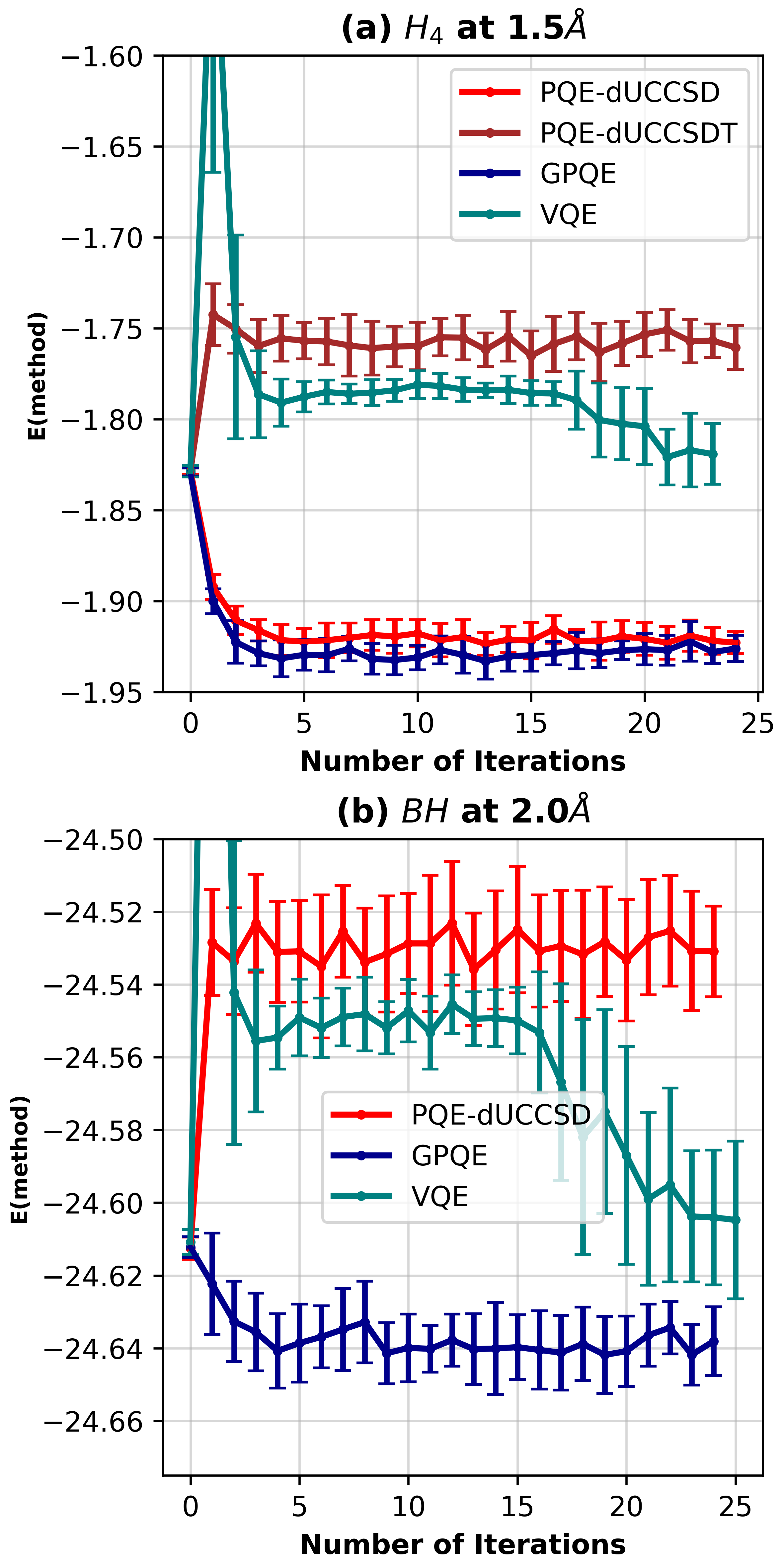}
    \caption{\textbf{Energy is plotted at each iteration under depolarizing noise channel. This energy is averaged over 20 independent runs and error bar denotes the standard deviation.}}\label{fig:de_noise}
  \end{figure}

\subsection{Numerical Simulation with a Depolarizing Noise Channel: Superior Noise Resilience of GPQE over VQE}

While our GPQE approach opens up a new direction to treat generalized 
operators in the projective framework that leads to less
quantum resource utilization over allied PQE family of methods,
such a strategy often leads to nearly identical results when
the parameters are variationally optimized in the noiseless 
VQE framework (Fig.\ref{fig:surface}). Thus the absolute superiority of our 
approach over VQE with an equivalently 
parametrized ansatz warrants further numerical validation 
under the realistic noisy environment.  
To this end, we simulated the ansatz within both VQE and 
GPQE frameworks with one- and two-qubit depolarizing noise
channels,
which can mathematically model the primary source of errors in NISQ hardware.
We applied one- and two-qubit depolarizing 
errors of the order of $10^{-5}$ and $10^{-4}$, respectively, 
and simulated the linear $H_4$ model at $R_{H-H}=1.5$\AA\hspace{1mm} and $BH$ at $R_{B-H}=2$\AA \hspace{1mm}
by averaging over 20 independent circuit runs.

Dramatically, as shown in Fig.\ref{fig:de_noise}, even under
such low intensity of noise, the optimized GPQE energy is 
significantly lower than the optimized VQE energy. 
Interestingly, the GPQE energy optimization trajectory 
exhibits a noticeably steeper convergence dip,
in both the cases as compared to VQE. Consequently, in both Fig.\ref{fig:de_noise}(a) and (b), GPQE shows consistently
better average energy predictions with much less standard deviation than its VQE counterpart.
Another interesting observation is that, in Fig.\ref{fig:de_noise}(a) the energy of the PQE-dUCCSDT ansatz is much higher than that 
of both PQE-dUCCSD and GPQE, as expected, owing to the 
larger number of two-qubit gates in the dUCCSDT ansatz.
Due to this the PQE-dUCCSDT trajectory for 
Fig.\ref{fig:de_noise}(b) is not explicitly
shown as it lies way above the scale of the plot along y-axis.
In Fig. \ref{fig:de_noise}(a) we observe that PQE-dUCCSD energy 
profile lies close (although slightly higher, 
consistently) to that of our method, contrary to the noiseless behaviour (shown in Fig.\ref{fig:surface}).
Such an improved performance of PQE-dUCCSD (compared to 
our approach and over PQE-dUCCSDT in particular) is 
artifactual due to 
being less perturbed by noise and is not due to the capture 
of adequate correlation effects.
However, in case of Fig.\ref{fig:de_noise}(b) as the number of CNOT gates
are substantially high for PQE-dUCCSD, huge accumulation of noise affects the energy computations in a way that it does not even show
any sign of convergence as can be seen from the
relative behaviour of the trajectories.
Thus while we may expect to
have improved fault-tolerance of quantum device over next
few decades, such noisy energy profile would gradually approach 
to their corresponding noise-free limits as given by 
Fig. \ref{fig:surface} for which right numerical trend is 
solely dictated by 
the amount of correlation captured by the theory. In this
context, our method shows the best compromise between accuracy
and resource efficiency in its noiseless limit while maintaining requisite
accuracy in present day's noisy architecture. More importantly, in the current noisy
architecture, the improved performance of GPQE over 
VQE convincingly demonstrates the former's robustness
and superior noise resilience.

\section{Conclusions and future outlooks:}
\label{concl}

In this manuscript, we have developed a methodology, GPQE, for determining the amplitudes 
corresponding to the generalized operators within the projective quantum eigensolver
framework. Given the existence of the vacuum annihilating condition of the 
generalized operators when acted on the HF reference, the 
conventional Hilbert-space projection based PQE required to 
deal with only excitation type of operators which rapidly 
incurs high computational expenses due to the inclusion of 
high-rank excitations to achieve chemical accuracy for 
strongly correlated systems. In GPQE, we bypassed 
this issue by projecting the effective Hamiltonian against 
a set of contracted excited determinants, the number of which
precisely equals the number of generalized unknown 
parameters. The residue determining equations are shown to 
be easily implemented in quantum computers in terms of the 
sum of diagonal expectation values, somewhat 
akin to the traditional formulation. 

The solution of the generalized operators with GPQE 
opens up a new research direction in which one 
can maintain minimal circuit depth to simulate electronic 
strong correlation within PQE framework while concurrently 
retaining its high noise resilience in NISQ 
architecture. The advantage of this development is 
demonstrated under ideal 
environment where GPQE with a parametrized double unitary 
ansatz, containing both excitation and
generalized operators, and is shown to be as accurate as 
dUCCSDT-PQE while the former requires orders of magnitude less quantum resources. This is also shown to be 
more accurate than dUCCSD-PQE and dUCCSDT-PQE when simulated under a Gaussian noise model.
More importantly, when simulated under synthetic depolarising noise channel,
GPQE is shown to be more resilient than dUCCSD-PQE, dUCCSDT-PQE and VQE (with 
similar ansatz), demonstrating its potential as an alternative to VQE to simulate 
atoms and molecules with chemistry inspired ansatz in near-term quantum computers.

A CNOT-efficient implementation of GPQE would be a good simulation protocol towards
accurate determination of molecular energetics under NISQ devices. Furthermore, one may
integrate various error mitigation schemes for its practical realization.

\section{Acknowledgement}
DM thanks Prime Minister's Research Fellowship (PMRF), Government of India for his research fellowship. CP acknowledges University Grants Commission (UGC) and DH thanks Industrial Research and Consultancy Center (IRCC), IIT Bombay for their research fellowships.

\section*{Author Declarations}
\subsection*{Conflict of Interest:}
The authors have no conflict of interests to disclose.

\subsection*{DATA AVAILABILITY}
The numerical data that support the findings of this study are
available from the corresponding author upon 
reasonable request.

\subsection*{REFERENCES}
\bibliography{literature}
\end{document}